\documentclass[aps,prl,reprint,twocolumn,amsmath,amssymb,amsfonts]{revtex4-1}

\AtBeginDocument{\usepackage{booktabs}}               
\makeatletter
\g@addto@macro\bfseries{\boldmath}
\makeatother

\usepackage[varg]{txfonts}
\usepackage[T1]{fontenc}
\usepackage[utf8]{inputenc}

\usepackage{hyperref}
\usepackage{amsmath}
\usepackage{color}
\usepackage{graphicx}
\usepackage[percent]{overpic}
\usepackage{mathrsfs}
\usepackage{bm}
\usepackage{braket}
\usepackage[nolist,nohyperlinks]{acronym}

\usepackage{comment}

\newcommand{\subref}[2]{\ref{#1}\hyperref[#1]{#2}}

\newcommand{\avg}[1]{\braket{#1}}

\newcommand{\del}{\partial}

\newcommand{\h}[1]{{#1}^{\dagger}}
 
\newcommand{\cc}[1]{{#1}^{*}}

\newcommand{\im}{{\rm Im}}

\newcommand{\Tr}{{\rm Tr}}
\newcommand{\hc}{{\rm h.c.}}

\renewcommand{\vec}[1]{\boldsymbol{#1}}
\newcommand{\mat}[1]{\vec{#1}}
\newcommand{\trp}[1]{{#1}^{\intercal}}
\newcommand{\vhat}[1]{\vec{\hat{#1}}}

\definecolor{cred}{RGB}{228,26,28}
\definecolor{cblue}{RGB}{55,126,184}
\definecolor{cdblue}{RGB}{40,96,139}
\definecolor{clblue}{RGB}{205,223,237}
\definecolor{cgreen}{RGB}{77,175,74}
\definecolor{cgray}{RGB}{150,150,150}
\definecolor{clgray}{RGB}{200,200,200}
\definecolor{cpurple}{RGB}{152,78,163}
\definecolor{corange}{RGB}{255,127,0}
\definecolor{cgold}{RGB}{230,171,2}

\definecolor{cL}{RGB}{204,71,120}
\definecolor{cG}{RGB}{204,204,204}

\hypersetup{colorlinks=true,linkcolor=cL,citecolor=cL,urlcolor=cL}

\newcommand{\shortsec}[1]{\emph{#1:}}

\newacro{DM}[DM]{{Dzyaloshinskii-Moriya}}

\begin{document}

\title{Non-Hermitian topology of spontaneous magnon decay}
\author{Paul A. McClarty}
\author{Jeffrey G. Rau} 
\affiliation{Max-Planck-Institut f\"ur Physik komplexer Systeme, 01187 Dresden, Germany}

\begin{abstract}
Spontaneous magnon decay is a generic feature of the magnetic excitations of anisotropic magnets and isotropic magnets with non-collinear order. In this paper, we argue that the effect of interactions on one-magnon states can, under many circumstances, be treated in terms of an effective, energy independent, non-Hermitian Hamiltonian for the magnons. In the vicinity of Dirac or Weyl touching points, we show that the spectral function has a characteristic anisotropy arising from topologically protected exceptional points or lines in the non-Hermitian spectrum. Such features can, in principle, be detected using inelastic neutron scattering or other spectroscopic probes. We illustrate this physics through a concrete example:  a honeycomb ferromagnet with Dzyaloshinskii-Moriya exchange. We perform interacting spin wave calculations of the structure factor and spectral function of this model, showing good agreement with results from a simple effective non-Hermitian model for the splitting of the Dirac point. Finally, we argue that the zoo of known topological protected magnon band structures may serve as a nearly ideal platform for realizing and exploring non-Hermitian physics in solid-state systems.
\end{abstract}

\date{\today}

\maketitle

Understanding the role of topology in condensed matter physics has been one of the principal goals of a generation of physicists~\cite{haldane2017nobel,bernevig2013topological,hasan2010colloquium,burkov2016topological,chiu2016classification,bansil2016colloquium,yan2017topological,armitage2018weyl}. 
The topic has been profoundly fruitful in recent years on both the theoretical and experimental fronts and its current significance is reflected in its ubiquity, driving progress nearly all subfields of condensed matter physics. One relatively recent direction comes from progress in fabrication of photonic devices with engineered levels of loss or dissipation of photons from the system~\cite{lu2014topologicalphotonics,feng2017non}.
This provides an experimentally realizable context where \emph{non-Hermitian} terms in the effective tight-binding Hamiltonian can be important. 
Such terms can lead to new topological band structures that are distinct from their Hermitian counterparts. As in the conventional case, there is a wide variety of such phases, depending on symmetry and dimensionality~\cite{rudner2009topological,ozawa2018topological,esaki2011edge,liang2013topological,zeuner2015observation,leykam2017edge,ke2017topological,kozii2017non,shen2018topological,zhou2018observation,papaj2018nodal,gong2018topological,xiong2018does,lieu2018topological,yao2018non,kunst2018biorthogonal,alvarez2018topological,zyuzin2018flat,lee2018anatomy,kawabata2019topological,yoshida2018,yoshida2019,bergholtz2019weyl}. 

Whereas in photonic systems dissipation comes about because the substrate used to channel light can be lossy, effective non-Hermitian terms can also arise in closed quantum systems~\cite{nonhermitian2009rotter,moiseyev2011non,cao2015non}. By isolating a subsystem consisting of the single quasi-particle states one may trace out the remaining (reservoir) degrees of freedom to arrive at an effective non-Hermitian momentum and frequency dependent Hamiltonian $H_{\rm eff}(\boldsymbol{k},\omega)=H(\vec{k})+\Sigma(\vec{k},\omega)$ where $H(\vec{k})$ is the Hermitian part and the non-Hermitian part enters through the self-energy $\Sigma(\vec{k},\omega)$. This effective Hamiltonian describes propagating quasi-particle excitations with finite lifetime and potentially nontrivial non-Hermitian topology. Various such cases have been discussed in the literature in the context of topological matter, where the self-energy is generated by processes such as electron-electron, electron-impurity, finite temperature electron-photon scattering or by coupling to finite temperature leads~\cite{kozii2017non,yoshida2018,yoshida2019,bergholtz2019weyl}. 

\begin{figure}[tp]
    \centering
    \includegraphics[width=0.8\columnwidth]{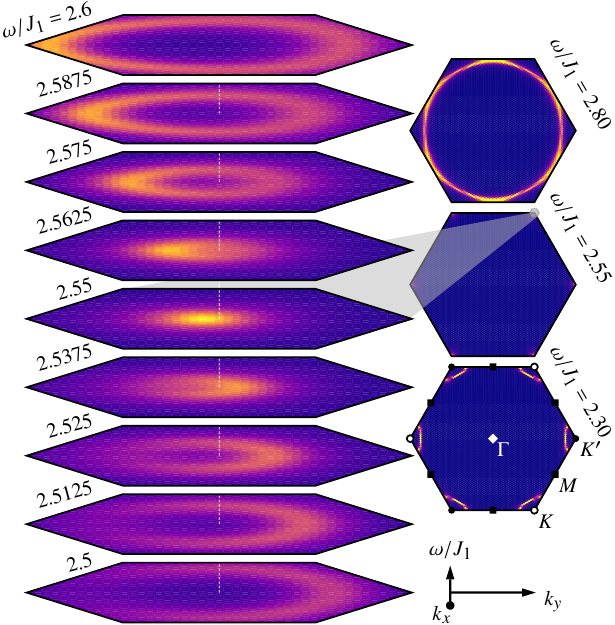}
    \caption{\label{fig:slices}
    Spectral function, $A(\vec{k},\omega)$, of the honeycomb ferromagnet with Dzyaloshinskii-Moriya interactions, see Eq.~(\ref{eq:model}), on constant energy slices as a function of wave-vector near the the $K$ point. The magnon lifetime is shortest in the lower band on one side of the $K$ point and, in the higher energy band, on the opposite side of $K$. Energy slices of the full zone are also shown.}
\end{figure}

In this paper, we propose that magnonic systems may be a nearly ideal platform to realize non-Hermitian topological states. Topological magnon states have been widely discussed in the Hermitian context~\cite{shindou2013topological,mook2014magnon,mook2014edge,romhanyi2015hall,chisnell2015topological,chernyshev2016damped,owerre2016first,mcclarty2017topological,owerre2017noncollinear,nakata2017magnonic,kitaevtopologicalmagnons2018}. One of the peculiarities of topological magnons compared to their electronic cousins is that they appear as \emph{excited} states and therefore they are subject to the non-universal effects of interactions~\cite{chernyshev2016damped,mcclarty2017topological,kitaevtopologicalmagnons2018}. Recent theoretical work has demonstrated that topological magnon edge states can be non-perturbatively sharp and observable in principle despite the presence of magnon-magnon interactions~\cite{kitaevtopologicalmagnons2018}. Nevertheless, intrinsic magnon decay coming from one-to-two magnon processes is generic in magnetically ordered systems, particularly those with some degree of anisotropic exchange. Such decay of otherwise sharp magnon states into a multi-magnon continuum supplies a natural separation into system and reservoir, and motivates the treatment of one-magnon states as a non-Hermitian Hamiltonian problem.

The simplest realization of this physics occurs when the non-interacting problem contains linear touching points. Without decay, it is known that the inelastic neutron scattering intensity exhibits a characteristic signature that manifests as a winding of the intensity around the touching point in constant energy slices~\cite{shivam2017neutron}. In the presence of non-Hermitian terms, linear touching points evolve into topologically protected exceptional points (in 2D) or lines (in 3D) that bound degenerate lines or surfaces in the real part of the energy~\cite{ozawa2018topological,esaki2011edge,liang2013topological,zeuner2015observation,leykam2017edge,ke2017topological,kozii2017non,shen2018topological,zhou2018observation,papaj2018nodal,gong2018topological,xiong2018does,lieu2018topological,yao2018non,kunst2018biorthogonal,alvarez2018topological,zyuzin2018flat,lee2018anatomy,kawabata2019topological}.  The presence of these features results in a universal anisotropy in the magnon lifetime around the touching point as shown in Fig.~\ref{fig:slices}. 

\shortsec{Spin wave theory} We first recall the notation and concepts of spin-wave theory. We consider a magnetically ordered system and expand about the classical limit using Holstein-Primakoff bosons~\cite{holstein1940,kittel1963quantum,auerbach1998interacting}. At leading order, the effective Hamiltonian takes the form
\begin{equation}
  \sum_{\vec{k}} \sum_{\alpha\beta} \left[ A^{\alpha\beta}_{\vec{k}} \h{a}_{\vec{k}\alpha} a^{}_{\vec{k}\beta}+
    \frac{1}{2}\left(B^{\alpha\beta}_{\vec{k}} \h{a}_{\vec{k}\alpha} \h{a}_{-\vec{k}\beta}+\hc\right)
    \right],
\end{equation}
where $a_{\vec{k}\alpha}$ is the (Fourier-transformed) Holstein-Primakoff boson with wave-vector $\vec{k}$ on sublattice $\alpha$ of the (magnetic) unit cell. The matrices $\mat{A}_{\vec{k}}$ and $\mat{B}_{\vec{k}}$ depend on the classical ordering pattern and the details of the exchange model. The linear spin-wave spectrum is determined by the eigenvalues of the Bogoliubov dispersion matrix~\cite{blaizot1986quantum}
\begin{equation}
  \mat{\sigma}_3 \mat{M}_{\vec{k}} \equiv
  \left(
    \begin{array}{cc}
      \mat{A}_{\vec{k}} & \mat{B}_{\vec{k}} \\
      -\cc{\mat{B}}_{-\vec{k}} & -\cc{\mat{A}}_{-\vec{k}} \\
    \end{array}
  \right),
\end{equation}
where $\mat{\sigma}_3 \equiv {\rm diag}(+\mat{1},-\mat{1})$ is a block Pauli matrix. 
While this generalized eigenvalue problem is not Hermitian, the matrix $\mat{M}_{\vec{k}}$ \emph{is} Hermitian, and satisfies $\mat{M}_{\vec{k}} = \h{\mat{M}}_{\vec{k}} = \mat{\sigma}_1 \trp{\mat{M}}_{-\vec{k}}\mat{\sigma}_1$.
The spin-wave modes obtained at leading order are thus stable quasi-particles, with infinite lifetime.

Going to higher order, one must consider magnon-magnon interactions. The effects of these interactions are encoded in the (retarded) self-energy, $\mat{\Sigma}(\vec{k},\omega)$, which can be computed perturbatively~\cite{negele1988quantum,blaizot1986quantum} in the limit $1/S \rightarrow 0$~\cite{dyson1956rmp,harris1971,zhitomirsky2013}. This can be defined via the (retarded) magnon Green's function as
\begin{equation}
  \label{eq:green}
  \mat{G}(\vec{k},\omega) \equiv \left[
    (\omega + i0^+)\mat{\sigma}_3
    -\left(\mat{M}_{\vec{k}}  + \mat{\Sigma}(\vec{k},\omega)\right)
    \right]^{-1}.
\end{equation}
Such interactions appear first at next-to-leading order, inducing both renormalization of the spin-wave spectrum and the possibility of spontaneous magnon decay.

\shortsec{Non-Hermitian Magnon Hamiltonians} To relate the magnon self-energy to an effective non-Hermitian band structure, we consider the perturbative limit where the self-energy is small, which can be reached, for example, as $S \rightarrow \infty$ or in very large magnetic fields. In this limit the one-magnon excitations are then characterized by a set of \emph{quasi}-normal modes that are defined by poles of the retarded Green's function~\cite{chernyshev2009}
\begin{equation}
\det\left((E_{\vec{k}}-i\Gamma_{\vec{k}} + i0^+)\mat{\sigma}_3
    -\left[\mat{M}_{\vec{k}}  + \mat{\Sigma}(\vec{k},E_{\vec{k}}+i\Gamma_{\vec{k}})\right]
    \right) = 0,
\end{equation}
where $E_{\vec{k}}$ is the energy of the quasi-normal mode and $\Gamma_{\vec{k}}$ is its inverse lifetime.
Away from the edges of the two-magnon continuum~\cite{zhitomirsky2013}, singular features in the two-magnon density of states, or degeneracies, one can solve this equation perturbatively~\cite{zhitomirsky2013}
\begin{equation}
    E_{\vec{k}\alpha}-i\Gamma_{\vec{k}\alpha} = \epsilon_{\vec{k}\alpha} 
    + \h{\vec{V}}_{\vec{k}\alpha} \mat{\sigma}_3\mat{\Sigma}(\vec{k},\epsilon_{\vec{k}\alpha})
    {\vec{V}}^{}_{\vec{k}\alpha},
\end{equation}
where $\epsilon_{\vec{k}\alpha}$ and $\vec{V}_{\vec{k}\alpha}$ are the non-interacting magnon energies and eigenvectors for band $\alpha$.

Now suppose the non-interacting magnon bands of interest lie within an energy window such that the frequency dependence of the self-energy can be neglected. This naturally occurs at band touchings, such as Dirac or Weyl points, nodal lines or surfaces or even entirely flat bands. If we denote the center of this energy window as $\omega_0$ then, in each case, we can expand the self-energy as
\begin{equation}
\mat{\Sigma}(\vec{k},\omega) \approx \mat{\Sigma}(\vec{k},\omega_0)
+(\omega-\omega_0)\del_{\omega}\mat{\Sigma}(\vec{k},\omega_0) + \cdots.
\end{equation}
Keeping only the first piece, the non-linear eigenvalue equation for the quasi-normal modes then becomes linear~\footnote{The second piece can also be retained without rendering the generalized eigenvalue problem non-linear. One then has something analogous to a ``quasi-particle weight renormalization'', $[\mat{1} - \mat{\sigma}_3\del_{\omega} \mat{\Sigma}(\vec{k},\omega_0)]^{-1}$ multiplying the effective dispersion matrix if we track the energies relative to $\omega_0$.}, with the effective
\emph{non-Hermitian} dispersion matrix
\begin{equation}
    \mat{M}^{\rm eff}_{\vec{k}} \equiv 
        \left[\mat{M}_{\vec{k}} + \mat{\Sigma}'(\vec{k},\omega_0)\right]+\mat{\Sigma}''(\vec{k},\omega_0),
\end{equation}
where we have separated the Hermitian ($\mat{\Sigma}'$) and anti-Hermitian ($\mat{\Sigma}''$) parts explicitly. Thus within this energy window, we have an explicitly non-Hermitian band structure and thus can explore a range of fundamentally non-Hermitian phenomena. We note that this effective non-Hermitian problem is not entirely general, as causality enforces the constraint that imaginary parts of the quasi-normal modes must always be \emph{negative}.

\shortsec{Honeycomb ferromagnet} We now consider a simple illustrative example. We consider a spin-1/2 ferromagnet on the honeycomb lattice in the presence of a second-neighbor in-plane \ac{DM} interaction
\begin{equation}
\label{eq:model}
    H = -\sum_{n=1}^3\sum_{\avg{ij}_n} J_n\vec{S}_i\cdot\vec{S}_j + 
    D \sum_{\avg{ij}_2} (-1)^i \vhat{r}_{ij} \cdot (\vec{S}_i\times \vec{S}_j),
\end{equation}
where $\vhat{r}_{ij}$ is the direction from site $i$ to $j$, and $(-1)^i$ is a staggered sublattice sign (to preserve inversion symmetry).
To enhance the effects of magnon interactions, we include small second and third neighbour ferromagnetic couplings with $J_2/J_1 = 0.2$ and $J_3/J_1=0.1$~\footnote{We note that an out-of-plane  \ac{DM} interaction can also be included, though this does not strongly affect our results.}. 
The ground state of this model is a collinear ferromagnet, with the moment direction arbitrary; to preserve the two-fold symmetry, we perform our semi-classical expansion about a ground state polarized along $\vhat{x}$. Similar to graphene, at leading order the spectrum hosts symmetry protected Dirac touching points at the corners of the Brillouin zone (wave-vectors $K$ and $-K$) located at $\omega_0 = 3(J_1 + 3J_2 + J_3)/2$, independent of the value of $D$.

\begin{figure}[tp]
    \centering
    \includegraphics[width=\columnwidth]{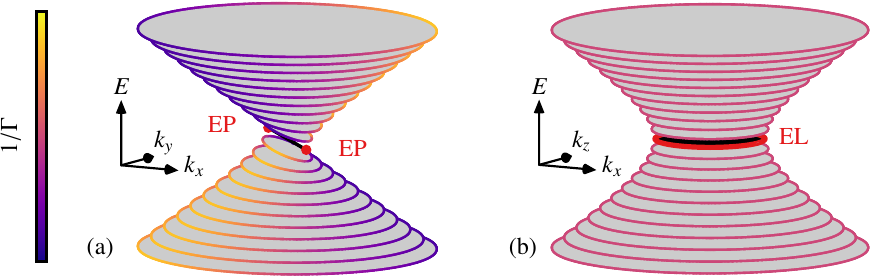}\null
    \vspace{0.5cm}
    \includegraphics[width=\columnwidth]{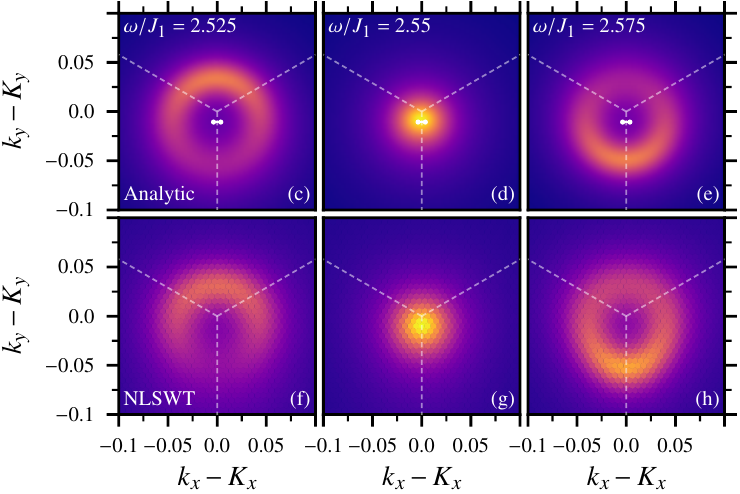}
    \caption{\label{fig:spectral}
    (a-b)  Constant slices of the real part of the quasi-normal modes,  $E_{\vec{k}}$, as a function of momentum for the touching point in (a) 2D, showing the arc joining exceptional points (EPs) and (b) 3D showing the surface bounded by exceptional lines (ELs). The inverse of the imaginary part of the modes, $1/\Gamma_{\vec{k}}$, is also shown.
    (c-h) Comparison of the spectral function, $A(\vec{k},\omega)$, evaluated for $D/J_1 = 0.125$ near the Dirac touching for the (c-e) analytic effective model (with the arc and exceptional points indicated) using $\mat{\Sigma}^{-+}({K},\omega_0)$ [Eq.~(\ref{eqn:sigma})] and (f-h) the result from non-linear spin-wave theory (NLSWT), with a full solution of Dyson's equation.
    }
\end{figure}

While it does not strongly affect the linear spin-wave spectrum, the \ac{DM} interaction is important at higher-order, generating spontaneous magnon decay. Indeed, from simple kinematics one can see that the Dirac touching points sit well within the two-magnon continuum and thus are liable to decay spontaneously once the \ac{DM} interaction is turned on. Further, when the polarization is along $\vhat{x}$ these touchings are protected by symmetry so, while their position shifts, they are not lifted by the interaction-induced Hermitian perturbations generated by the \ac{DM} exchange. This model thus affords us an ideal setting to observe the effect of non-Hermitian perturbations on the Dirac touching points, with the \ac{DM} interaction serving as a tuning parameter to control their strength~\footnote{The strength of \ac{DM} also tunes away from the ferromagnetic limit where the one-magnon excitation spectrum is \emph{exact}. As $D/J_1 \rightarrow 0$, spin-wave theory is thus controlled at arbitrary spin, without an appeal to the classical $S \rightarrow \infty$ limit.}.

\shortsec{Two-band model} The effect of non-Hermitian perturbations to a Dirac touching point has been thoroughly studied~\cite{shen2018topological,kozii2017non,papaj2018nodal,leykam2017edge}, so we only briefly review the relevant theory. One can write an effective two-band model near the (Hermitian) touching point as
\begin{equation}
    v\left( k_x\sigma_x + k_y\sigma_y\right)  -i\left(a_0 + \vec{a} \cdot \vec{\sigma}\right),
\end{equation}
where energy and momentum are measured relative to the nodal point, $\vec{\sigma}$ are the Pauli matrices and all parameters are real. Hermitian perturbations that do not lift the node can be absorbed into the definition of the energy and touching point, and thus have been discarded. The four parameters $a_0$, $\vec{a}$ represent the (leading) constant non-Hermitian perturbations. While $a_0$ is simply a constant width, the vector part has a more dramatic effect, splitting the Dirac point into a line that connects two (non-diagonalizable) exceptional points with a dispersion $\sim |\vec{k}|^{1/2}$, as illustrated in Fig.~\subref{fig:spectral}{(a)}. This arc is centered at the touching point, has length $|\vec{a}|/v$ and runs perpendicular to $\vec{a}_{\perp}=(a_x,a_y)$. This is a distinctly non-Hermitian topological phenomenon as can be seen by taking a closed path around one of the exceptional points. The two eigenvalues swap discontinuously upon crossing the arc and this is associated with a half-integer winding number \cite{shen2018topological}; this is visible in the linewidth, which changes discontinuously going through the arc, but smoothly going around it. These exceptional points are topologically protected, and must annihilate in pairs, thus surviving even in the presence of sufficiently small $\sigma_z$ mass terms in the Hamiltonian. We note that causality imposes the constraint that $a_0 \geq |\vec{a}|$, requiring some amount of diagonal broadening on top of the structure of the arc and exceptional points.

\shortsec{Comparison} Precisely this physics arises in the ferromagnetic honeycomb model given in Eq.~(\ref{eq:model}). The leading contribution to the self-energy comes from a single (bubble) diagram corresponding to decay and recombination of a single magnon, yielding only the normal self-energy
\begin{equation}
\label{eq:diagram}
    \Sigma^{-+}_{\alpha\beta}(\vec{k},\omega) = 
\frac{1}{N}\sum_{\vec{q}}\sum_{\rho\rho'}\frac{
\cc{[W^{\alpha,\rho\rho'}_{\vec{k},\vec{q}}]} W^{\beta,\rho\rho'}_{\vec{k},\vec{q}}}
{\omega- \epsilon_{\vec{q}\alpha} - \epsilon_{\vec{k}-\vec{q},\beta}+i0^+},
\end{equation}
where $\alpha,\beta$ are sublattice indices, $\rho,\rho'$ are band indices and $W^{\alpha,\rho\rho'}_{\vec{k},\vec{q}}$ is a form-factor~\cite{zhitomirsky2013}. This can be expanded about the Dirac touching at energy $\omega_0$ and at wave-vector ${K}$
to yield an effective non-Hermitian perturbation. For example, at $D/J_1~=~0.125$ this yields 
\begin{equation}
    \Sigma^{-+}_{\alpha\beta}({K},\omega_0)\approx
    \frac{D^2}{J_1}\left(
    \begin{array}{cc}
         0.08-0.95i & -0.41+0.28i \\
         -0.41+0.28i  & 0.08-0.95i
    \end{array}
    \right)_{\alpha\beta},
    \label{eqn:sigma}
\end{equation}
where the associated wave-vector sum in Eq.~(\ref{eq:diagram}) has been evaluated numerically for a large, but finite system.
As required by symmetry, the mass-like terms $\propto \sigma_z$ are absent, and both the Hermitian and non-Hermitian parts are non-zero and comparable.

To see how the picture obtained from the two-band model is borne out, we compare the results of this effective non-Hermitian description with the results of more complete interacting spin-wave calculation that retains the energy dependence of the self energy and solves the full Dyson equation. 
This comparison is shown in Figs.~\subref{fig:spectral}{(c-h)} for the magnon spectral function $A(\vec{k},\omega) \equiv -\im{\Tr{[\mat{G}(\vec{k},\omega)]}}/\pi$ for $D/J_1=0.125$ for several energies near the (renormalized) band touching. One can see good agreement, showing that the effective low-energy theory is a good description, capturing the physics of the magnon-magnon interactions. One can see the characteristic anisotropy of the linewidth directly in the interacting spin-wave theory looking at constant energy cuts near the $K$ point, as shown in Figs.~\ref{fig:slices} and \subref{fig:spectral}{(c-h)}.

\begin{figure}[tp]
    \centering
    \includegraphics[width=\columnwidth]{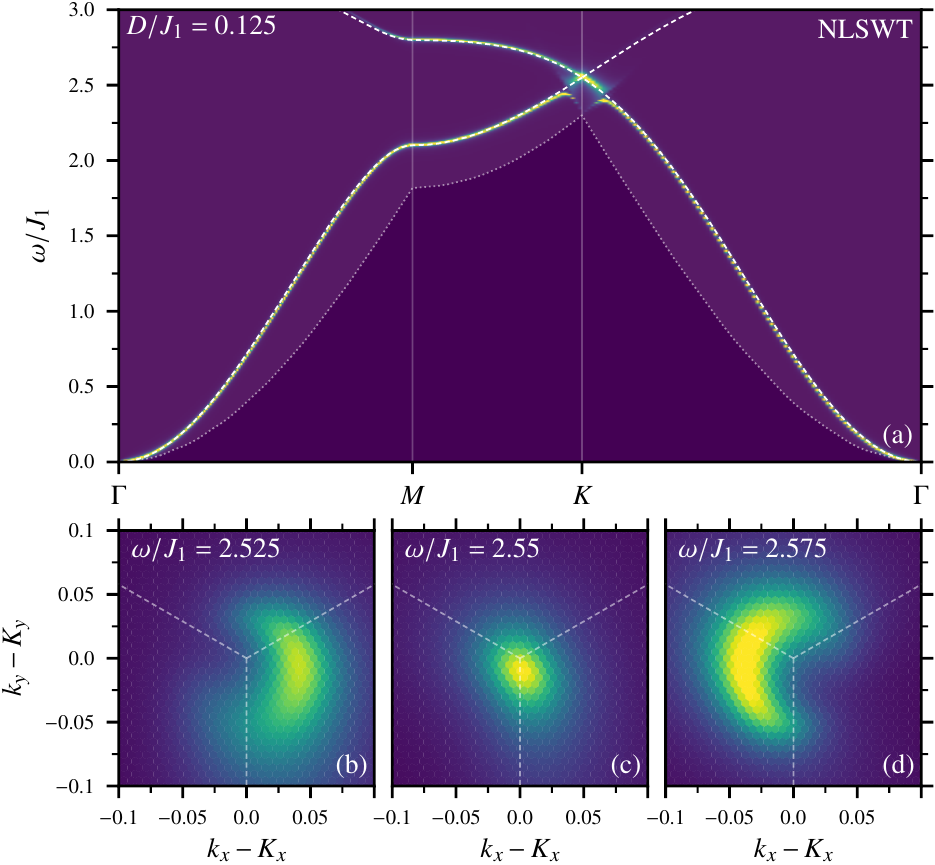}
    \caption{\label{fig:neutron}
    (a) The structure factor, $S(\vec{k},\omega)$, computed within NLSWT for $D/J_1=0.125$ along a path in momentum space. The linear spin wave dispersions (dashed) and bottom of the two magnon continuum (dotted) for $D/J_1=0$ are indicated.
    (b-d) Evolution of the neutron scattering intensity at a fixed $D/J_1=0.125$ as the energy is varied through the Dirac node, from (b) $\omega/J_1=2.525$ to (c) $\omega/J_1=2.55$ to (d) $\omega/J_1=2.575$.
    }
\end{figure}

\shortsec{Discussion and Outlook} The above example cleanly illustrates the non-Hermitian topological physics of magnons in the presence of Dirac band touchings in two dimensions. The existence of exceptional points and a line of degeneracies in the real part of the eigenvalue rests on the quasi-normal modes being well-defined which, in turn, relies on a meaningful separation of system and reservoir. At zero temperature, both the real and imaginary parts of the non-Hermitian effective Hamiltonian depend on the exchange alone so the above discussion is valid at weak coupling -- independent of the order of perturbation theory -- only potentially breaking down when interactions are strong. When the magnon-magnon coupling is intrinsically weak, temperature itself can provide an effective tuning parameter~\cite{thermodynamicdyson,pershoguba2018} for the non-Hermitian part of the effective magnon Hamiltonian. Thermal decay~\cite{pershoguba2018} can thus directly mirror the above discussion of spontaneous decay, providing another route to realizing this kind of non-Hermitian physics, so long as one is well below the ordering temperature.

The characteristic pattern of line broadening reported above is expected to be directly visible using inelastic neutron scattering in magnetic materials that exhibit Dirac or Weyl points within linear spin wave theory. We illustrate the dynamical structure factor, $S(\vec{k},\omega)$, as would be seen in inelastic neutron scattering~\cite{lovesey1984theory}, in Fig.~\ref{fig:neutron}. One can see the characteristic behavior of the broadening near the Dirac touching, as a function of both wave-vector and energy. In particular, while the intensity is modulated going away from the zone center~\cite{shivam2017neutron}, the modulation of the width follows the direction of the exceptional line. We note that similar anisotropies in the broadening are expected in the vicinity of the exceptional lines [Fig.~\subref{fig:spectral}{(b)}] around non-Hermitian Weyl points in three-dimensions. While observation of such features experimentally is potentially challenging given their small extent in wave-vector and energy, progress should be possible through careful studies using time-of-flight instruments or perhaps using specialized techniques such as NRSE \cite{TRISP,RESPECT}. 

There is an enormous variety of real magnetic materials that have been synthesized, many with the significant spin-orbit coupling or the non-collinear ground states needed for magnon decay to be relevant experimentally. Among such materials, in three-dimensions, the requirement of non-interacting spin wave spectra with Weyl touching points is expected not to be a particularly stringent condition~\cite{armitage2018weyl}, while in two-dimensions a discrete remnant symmetry is required~\cite{bernevig2013topological}.

A few possible examples of magnetic materials with linear touching points in the magnon spectrum that may also exhibit significant magnon interaction effects include the quasi-2D honeycomb materials CrBr$_3$ \cite{crbr3neutron,crbr3neutron2,pershoguba2018} and CrI$_3$ \cite{chentopologicalspinexcitations}, the 3D antiferromagnet Cu$_3$TeO$_6$ \cite{yao2018topological} and the possible Weyl magnon system Lu$_2$V$_2$O$_7$ \cite{mook2016weyl,mena2014lvo}. Another potentially interesting case is the kagome ferromagnet Cu$($1,3-bdc$)$ \cite{chisnell2015topological} that is thought to have significant anti-symmetric exchange couplings. In this case, the non-interacting magnon theory exhibits Chern bands. Chern bands are also well-defined in a non-Hermitian setting, with topologically protected edge states that are continuously connected to their Hermitian counterparts~\cite{shen2018topological}. 

Alongside the natural variety of magnets in nature, many magnets have ground states or excitations that are tuneable using experimentally accessible magnetic fields. This provides a possible tuning parameter that is not available in photonic or acoustic realizations of non-Hermitian topological states. A magnetic field will tend to cause a non-trivial evolution of the magnon spectrum as well as shifting the band center to higher energies. The latter effect can separate the one-magnon states from the multi-magnon continua thus supplying a mechanism to tune and ultimately switch off the non-Hermitian terms. Because the boundaries of the continua are sharp, the exceptional points may annihilate  discontinuously. 

The utility of field tunability can be illustrated vividly in our honeycomb ferromagnet example, where one finds that the Dirac point sits above a set of van-Hove singularities in the two-magnon density of states. Under application of a magnetic field, $h_x$, along the moment direction $\vhat{x}$, both the one- and two-magnon energies rigidly shift, but at different rates ($h_x/2$ and $h_x$ respectively). Thus by tuning the field the van-Hove singularity can be pushed closer to the Dirac touching, further enhancing the effect of magnon decay. Given the \ac{DM} interaction is typically subdominant in transition metal magnets~\cite{khomskii2014transition}, such as Lu$_2$V$_2$O$_7$~\cite{mena2014lvo} or the CrX$_3$ family~\cite{chentopologicalspinexcitations,crbr3neutron,crbr3neutron2}, a protocol such as this potentially presents a practical experimental route to controlling the effects of magnon decay, allowing full exploration of the non-Hermitian physics.

Aside from the experimental inference of exceptional points in magnon spectra using inelastic neutron scattering or terahertz spectroscopy, various interesting issues remain including, for example, the effect of non-Hermitian terms on magnon transport~\cite{chen2018hall} and the potential for novel measurements of topological physics in magnonic crystals \cite{shindou2013topological}, the role of magnon-phonon coupling as a separate dissipation mechanism, the possibility of introducing gain by optically pumping magnons, and bypassing the causality restrictions of interaction induced non-Hermitian terms. Another closely related avenue for the realization of non-Hermitian topology is in lattice vibrations via the presence of spontaneous \emph{phonon} decay~\cite{kosevich2006crystal,nozieres2018theory}. Another set of questions relates to physics beyond the approximation of the multi-magnon states as an \emph{incoherent} bath and possible interesting quantum effects arising from \emph{coherent} coupling between single and multi-magnon states.

{\it Acknowledgments} -- We thank J. Budich, A. Chernyshev, R. Moessner and M. Mourigal for useful discussions.

\bibliography{draft}

\end{document}